\documentclass[a4paper]{jpconf}
\usepackage{graphicx}
\usepackage{amssymb}
\usepackage[percent]{overpic}
\usepackage{cite}

\begin{document}
\title{J/$\psi$ and $\psi(2S)$ measurement in $p$+$p$ collisions at $\sqrt{s} =$ 200 and 500 GeV with the STAR experiment}

\author{Barbara Trzeciak$^1$ for the STAR Collaboration}

\address{$^1$Faculty of Nuclear Sciences and Physical Engineering, Czech Technical University in Prague, Brehova 7, 115 19 Praha 1, Czech Republic}

\ead{trzecbar@fjfi.cvut.cz}

\begin{abstract}
In this paper, results on the J/$\psi$ cross section and polarization measured via the dielectron decay channel at mid-rapidity in $p+p$ collisions at $\sqrt{s}$ = 200 and 500 GeV in the STAR experiment are discussed. Also, first measurements of the J/$\psi$ production as a function of the charged-particle multiplicity density and of $\psi(2S)$ to J/$\psi$ ratio at $\sqrt{s}$ = 500 GeV are reported. 
\end{abstract}

\section{Introduction}

J/$\psi$ and $\psi(2S)$ are bound states of charm ($c$) and anti-charm ($\overline{c}$) quarks. 
Final color-singlet charmonium physical state can be produced via either color-singlet (CS) or color-octet (CO) intermediate $c\overline{c}$ state.
Based on this, there are different models that try to describe $c\overline{c}$ pair production and its hadronization to the physical charmonium state, such as Color Singlet Model (CSM), Color Evaporation Model (CEM), or more sophisticated Non-Relativistic QCD (NRQCD) calculations.
However, the charmonium production mechanism in elementary particle collisions is not yet exactly known.
For many years measurements of the charmonium cross sections have been used to test different production models. While many models can describe relatively well the experimental data on the J/$\psi$ cross section in $p+p$ collisions \cite{Abelev:2009qaa, Adamczyk:2012ey, Adare:2011vq, PhysRevLett.79.572, Acosta:2004yw, Aad:2011sp, Khachatryan:2010yr, Aaij:2011jh}, they have different predictions for the J/$\psi$ polarization. Therefore, measurements of the J/$\psi$ polarization may allow to discriminate among different models and provide new insight into the J/$\psi$ production mechanism. Further constraints for models can be obtained by looking at the J/$\psi$ production vs relative charged-particle event multiplicity.

\section{J/$\psi$ cross-section measurements at $\sqrt{s} =$ 200 and 500 GeV}

\begin{figure}[ht]
\center
		\includegraphics[width=0.4\linewidth]{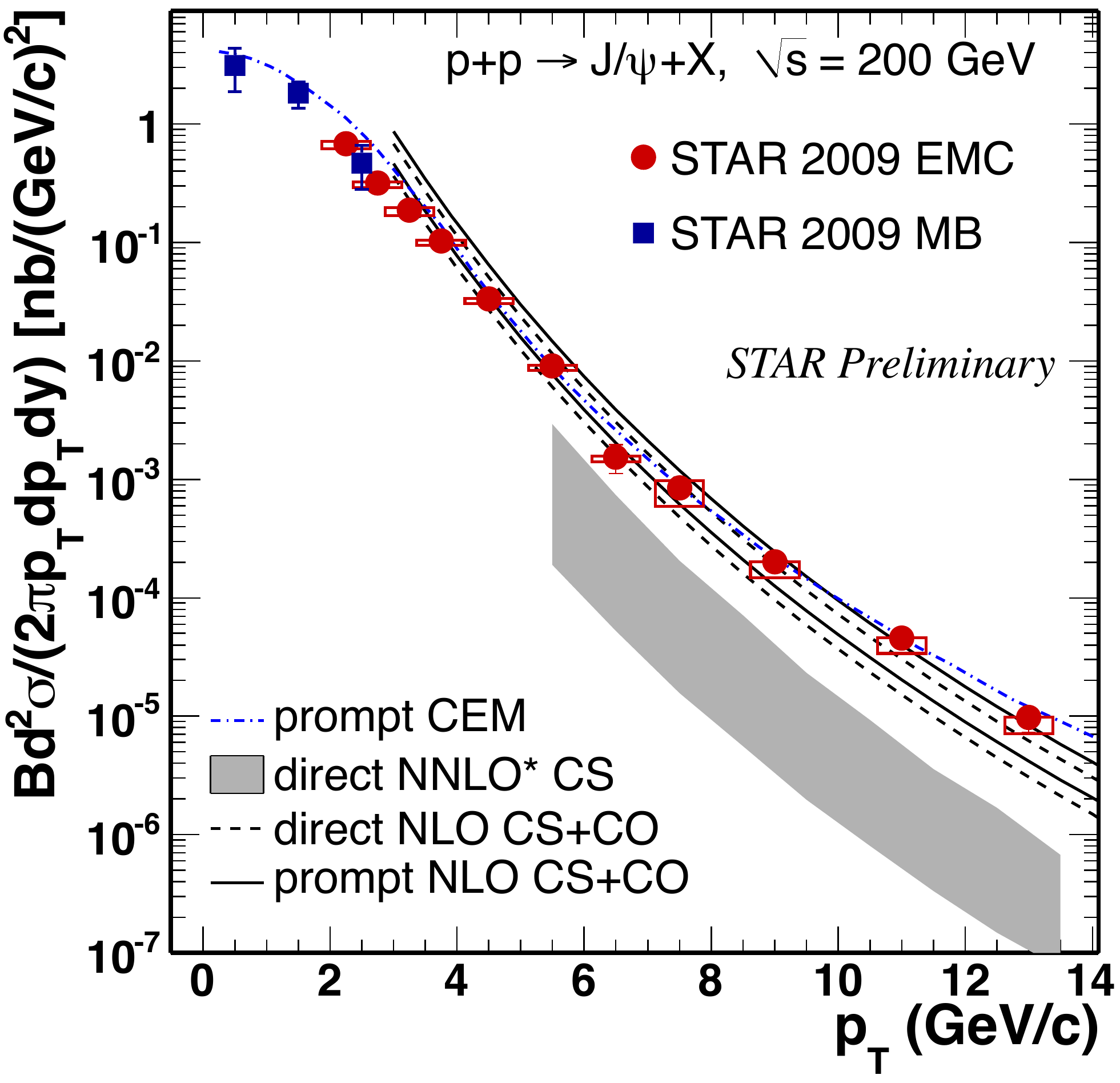}
		\includegraphics[width=0.4\textwidth]{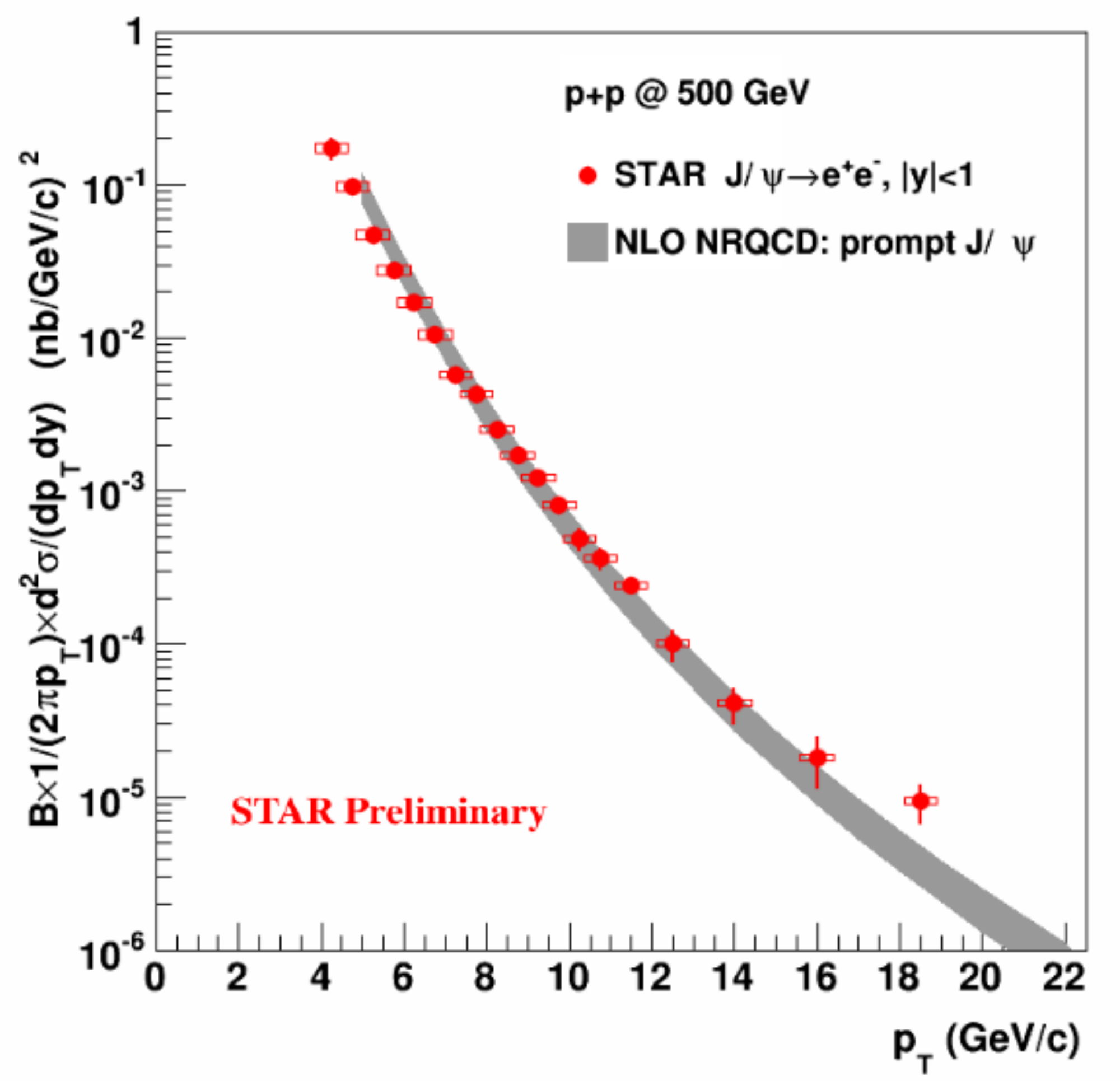}
		\caption{J/$\psi$ invariant cross section vs $p_{T}$ in $p$+$p$ collisions at mid-rapidity, left: at $\sqrt{s} =$ 200 GeV,  low~\cite{Kosarzewski:2012zz} and high $p_{T}$~\cite{Adamczyk:2012ey} results are shown as blue squares and red circles, respectively, and are compared to different model predictions~\cite{Frawley:2008kk,Ma:2010jj,Artoisenet:2008fc}, right: at $\sqrt{s} =$ 500 GeV, compared to NRQCD calculations~\cite{Ma:2010yw,Ma:2010jj,Shao:2014yta}.}
		\label{fig:JpsiXsection}
\end{figure}

STAR has measured inclusive J/$\psi$ $p_{T}$ spectra in $p+p$ collisions at $\sqrt{s} =$ 200 and 500 GeV via the dielectron decay channel ($B_{ee} =$ 5.9\%) at mid-rapidity ($\vert y \vert < $ 1). These results are compared to different model predictions in order to understand the J/$\psi$ production mechanism in elementary collisions.
The left panel of Fig.~\ref{fig:JpsiXsection} shows STAR low and high-$p_{T}$ measurements of J/$\psi$ $p_{T}$ spectra~\cite{Kosarzewski:2012zz, Adamczyk:2012ey} at $\sqrt{s} =$ 200 GeV compared to different model predictions: CEM ~\cite{Frawley:2008kk}, NNLO* CS~\cite{Artoisenet:2008fc} and NLO NRQCD calculations with both color-singlet and color-octet contributions (NLO CS+CO)~\cite{Ma:2010jj}. The NNLO* CS model prediction is for the direct J/$\psi$ production only, so it does not include contributions from $\psi(2S)$, $\chi_{C}$ and $B$-meson decays to J/$\psi$, which may account for as much as 40\% of the inclusive J/$\psi$ production. The right panel of Fig.~\ref{fig:JpsiXsection} shows high-$p_{T}$ results at $\sqrt{s} =$ 500 GeV compared to NLO NRQCD prediction for the prompt J/$\psi$ production~\cite{Ma:2010yw,Ma:2010jj,Shao:2014yta}. In both cases NLO NRQCD calculations describe the data well at a higher $p_{T}$ range, $p_{T} \gtrsim$ 4 GeV/$c$. The result at $\sqrt{s} =$ 200 GeV is also in a good agreement with the CEM prediction. 

Also, $x_{T}$ scaling of J/$\psi$ cross section is observed: $\frac{d^{2}\sigma}{2 \pi p_{T} dp_{T} dy} = g(x_{T})/(\sqrt{s})^{n}$, where $x_{T} = 2p_{T}/\sqrt{s}$, with $n =$ 5.6 $\pm$ 0.2 at mid-rapidity and $p_{T} >$ 5 GeV/$c$ for a wide range of colliding energies~\cite{Abelev:2009qaa}. At $\sqrt{s} =$ 500 GeV the same $x_{T}$ scaling of high-$p_{T}$ J/$\psi$ production is seen~\cite{Trzeciak:2014cma}.

\section{J/$\psi$ and $\psi(2S)$ measurements at $\sqrt{s} =$ 500 GeV}

\begin{figure}[ht]
		\includegraphics[width=0.45\linewidth]{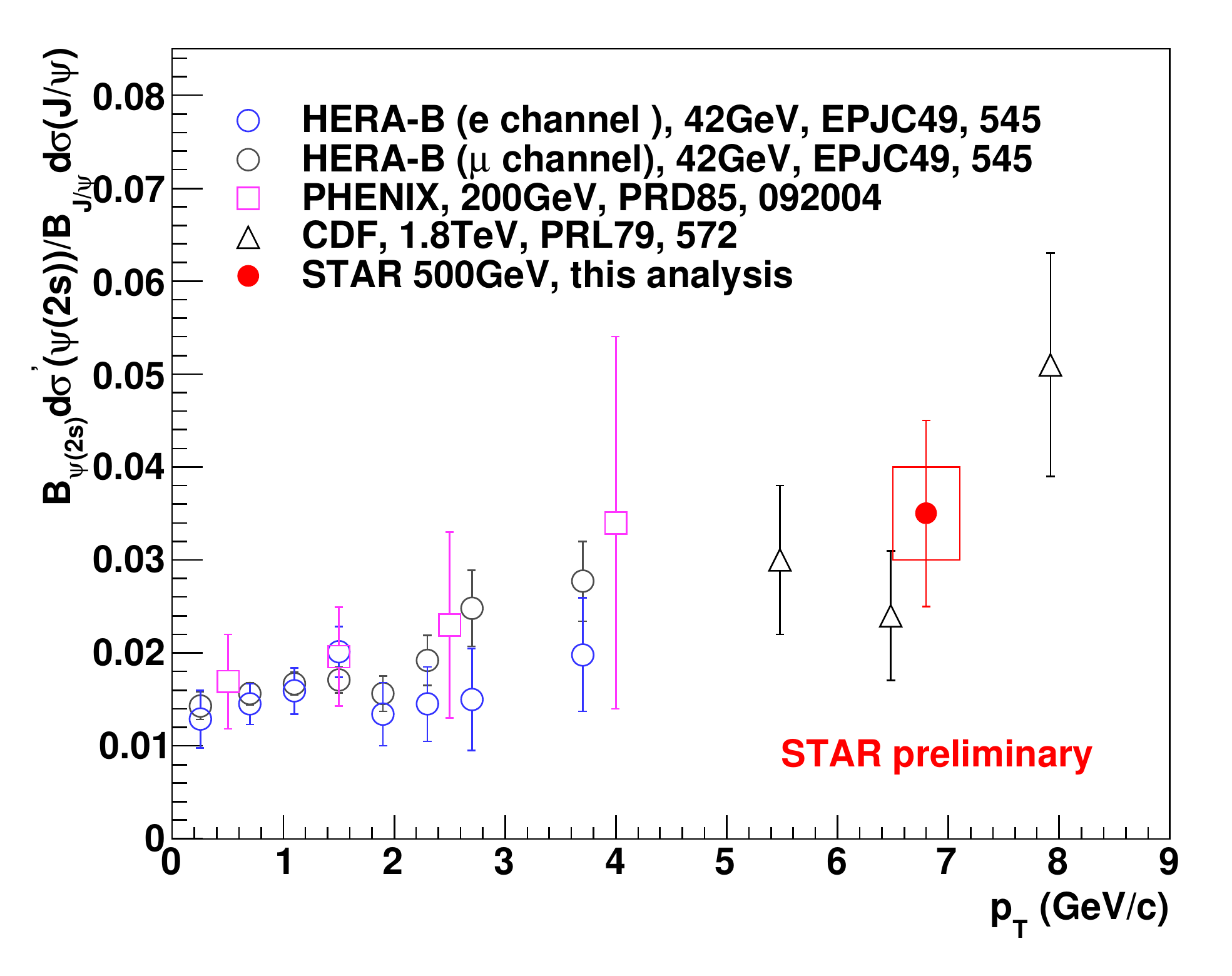}
		\includegraphics[width=0.42\linewidth]{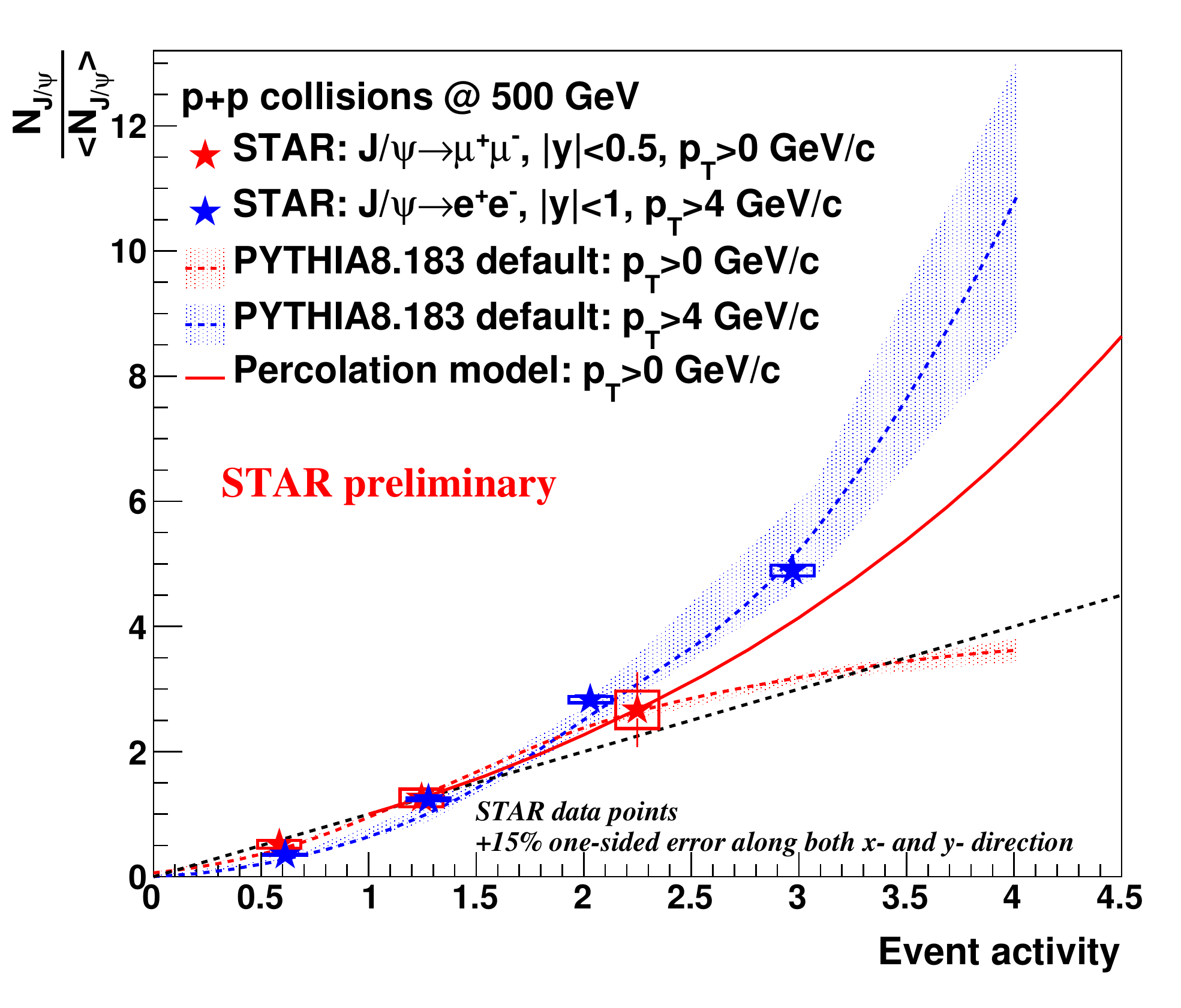}
		\caption{Left: ratio of $\psi(2S)$ to J/$\psi$ in $p+p$ collisions at $\sqrt{s} =$ 500 GeV from STAR (red circle) compared to results from other experiments at different energies. Right: Relative J/$\psi$ yield vs event activity compared to the default PYTHIA 8 tune and to the percolation model~\cite{Ferreiro:2012fb}.}
		\label{fig:Jpsi500}
\end{figure}

The left panel of Fig.~\ref{fig:Jpsi500} shows $\psi(2S) / J/\psi$ ratio from STAR (red full circle) compared to measurements of other experiments at different colliding energies, in $p+p$ and $p+$A collisions. The STAR data point is consistent with the observed trend, and no collision energy dependence of the $\psi(2S)$ to J$/\psi$ ratio is seen with the current precision.

STAR has also measured J/$\psi$ production vs event activity. The right panel of Fig.~\ref{fig:Jpsi500} shows relative J/$\psi$ yield as a function of relative charged particle event multiplicity. The measurement is done for integrated $p_{T}$ range in the dimuon decay channel and for $p_{T} >$ 4 GeV/$c$ in the dielectron decay channel, shown as red and blue points, respectively. We observe a strong correlation between the J/$\psi$ yield and event activity, and for $p_{T} >$ 4 GeV/$c$ there is a stronger than linear growth for higher multiplicities. The same behaviour has been observed for J/$\psi$ and open charm by the LHC experiments. There is also a hint of $p_{T}$ dependence of the correlation, but further analysis, especially in the dimuon decay channel for higher event activity bins is required in order to draw strong conclusions.
Results are also compared to the default PYTHIA 8 tune that includes multiple parton-parton interactions, for both analysed $p_{T}$ ranges, and to the percolation model~\cite{Ferreiro:2012fb} which is based on a string screening assumption. Both models can describe the observed increase of the J/$\psi$ relative yield. 

\section{J/$\psi$ polarization measurements at $\sqrt{s} =$ 200 and 500 GeV}
 
$J/\psi$ polarization is analyzed via the angular distribution of the decay electrons that is described by:
$\frac{d^{2}N}{d(\cos\theta)d\phi} \propto 1+\lambda_\theta \cos^2\theta + \lambda_\phi \sin^2\theta \cos2\phi + \lambda_{\theta\phi}\sin2\theta \cos\phi$, where $\theta$ and $\phi$ are polar and azimuthal angles, respectively; $\lambda_\theta$, $\lambda_\phi$ and $\lambda_{\theta\phi}$ are the angular decay coefficients. 
The first STAR measurement of the $J/\psi$ polarization was performed at $\sqrt{s} =$ 200 GeV at mid-rapidity and 2 $ < p_{T} <$ 6 GeV/$c$, in the helicity frame (HX)~\cite{Adamczyk:2013vjy}. Due to limited statistics only $\lambda_{\theta}$ polarization parameter, related to the polar anisotropy was extracted as a function of $p_{T}$. The result was compared to NRQCD calculations~\cite{Chung:2009xr} and the NLO$^{+}$ CSM prediction~\cite{Lansberg:2010vq}. A trend observed in the RHIC data is towards longitudinal polarization as $p_{T}$ increases and, within experimental and theoretical uncertainties, the result is consistent with the NLO$^{+}$ CSM prediction.

\begin{figure}[ht]
		\center
		\includegraphics[width=0.47\linewidth]{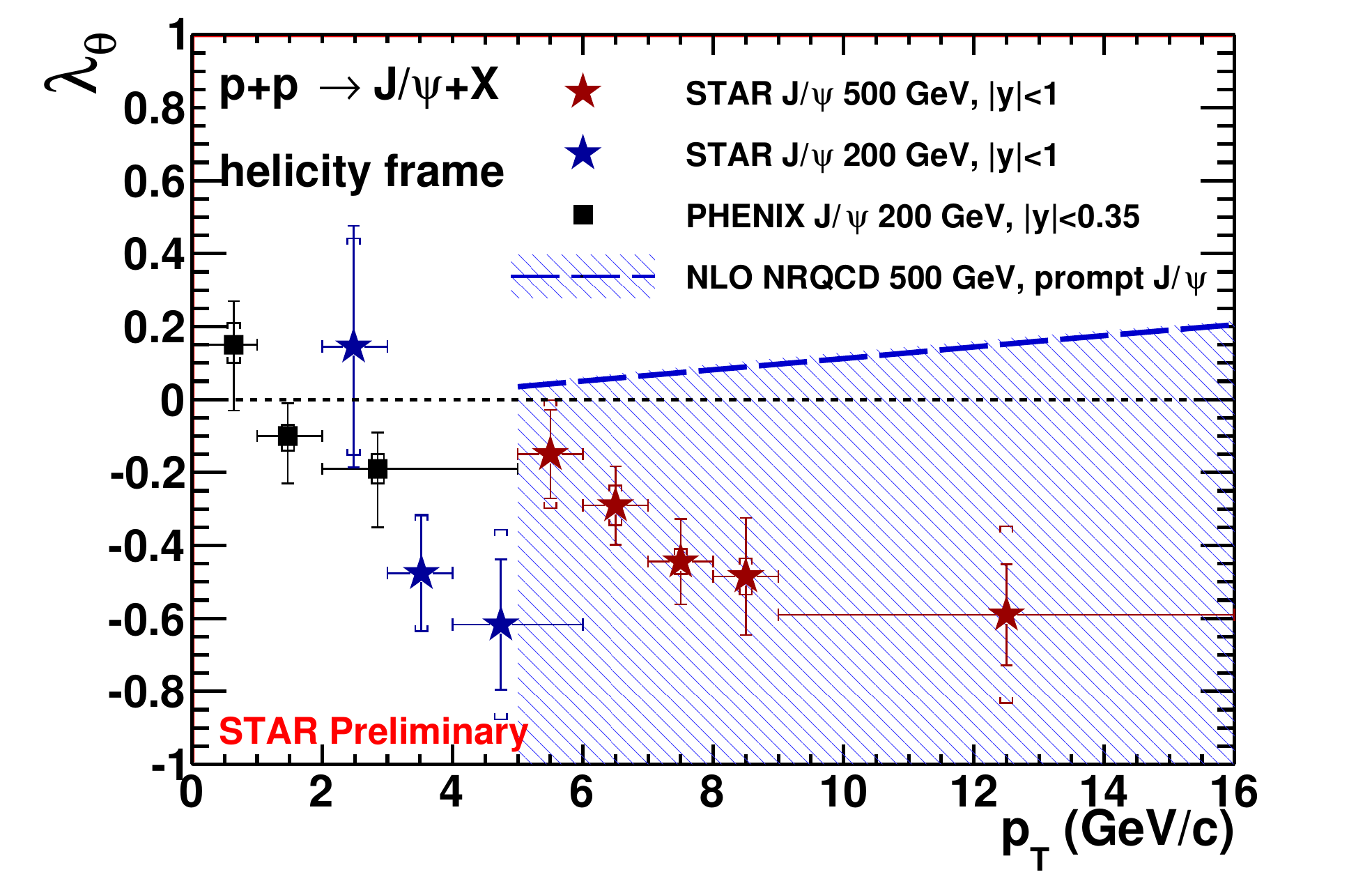}
		\hspace{20pt}
		\includegraphics[width=0.47\linewidth]{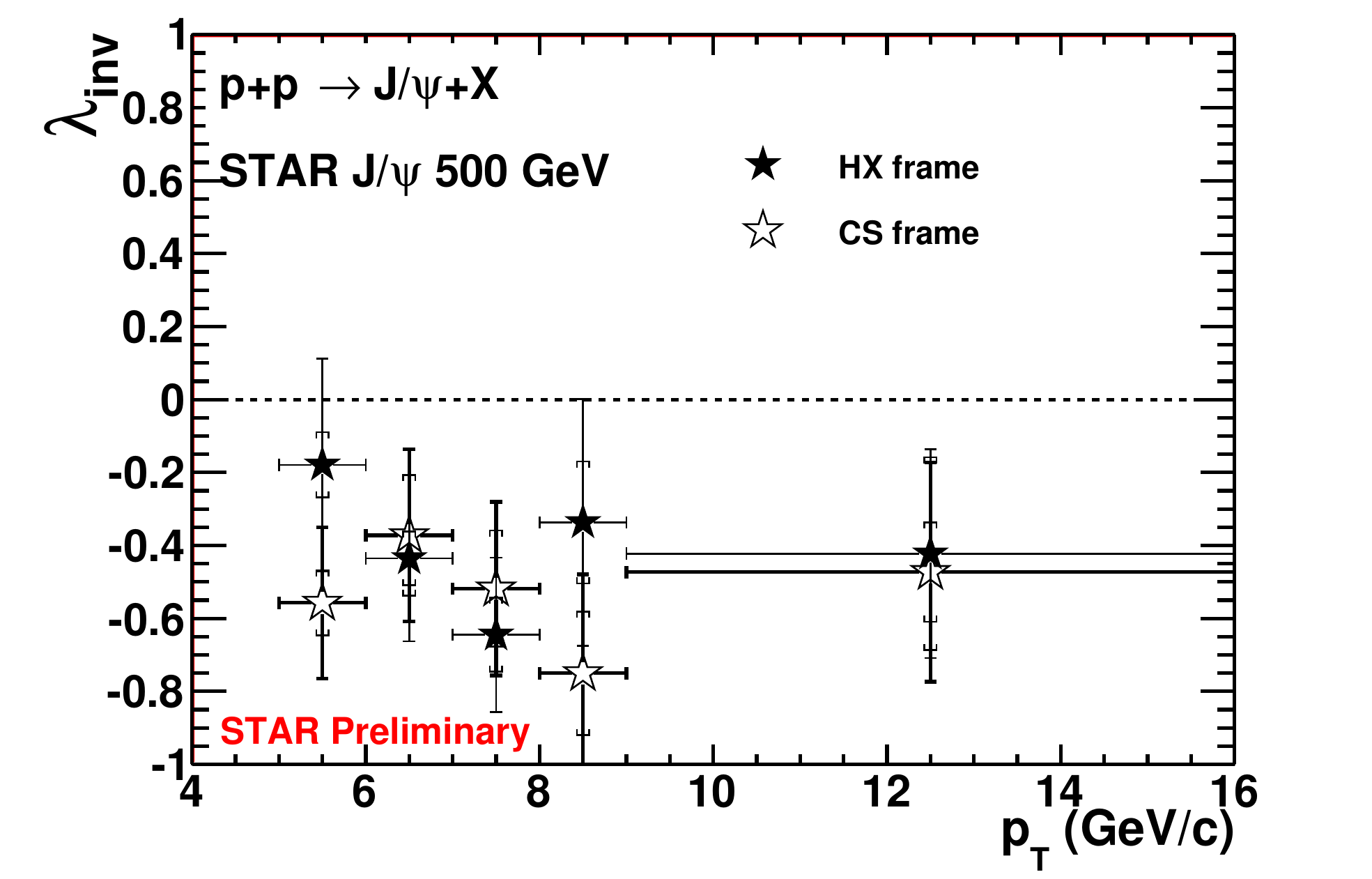}
		\caption{Left: J/$\psi$ polarization parameter $\lambda_{\theta}$ vs $p_{T}$ at $\sqrt{s} =$ 500 GeV at mid-rapidity compared to results at $\sqrt{s} =$ 200 GeV and to the NLO NRQCD calculations~\cite{Chao:2012iv,Shao:2012fs,Shao:2014fca,Shao:2014yta}. Right: $\lambda_{inv}$ parameters vs $p_{T}$ in the HX (full symbols) and CS frames (open symbols).}
		\label{fig:Jpsi500Polarization}
\end{figure}

The J/$\psi$ polarization measurements have been extended to a wider $p_{T}$ range of 5-16 GeV/$c$ using $p$+$p$ dataset at $\sqrt{s} =$ 500 GeV. Both $\lambda_\theta$ and $\lambda_\phi$ parameters have been extracted in the HX and the Collins-Soper (CS) frames. This has  also allowed to obtain the frame invariant parameter  $\lambda _{inv} = (\lambda_{\theta} + 3 \lambda_{\phi}) / (1- \lambda_{\phi})$ in both frames which is a good cross-check between measurements since any arbitrary choice of a reference frame should give the same value of the $\lambda_{inv}$~\cite{Faccioli:2010kd}. The left panel of Fig.~\ref{fig:Jpsi500Polarization} shows $\lambda_{\theta}$ in the HX frame as a function of $p_{T}$ at $\sqrt{s} =$ 500 GeV compared to results at $\sqrt{s} =$ 200 GeV and to a prediction of the NLO NRQCD~\cite{Chao:2012iv,Shao:2012fs,Shao:2014fca,Shao:2014yta}. We observe similar trend of $\lambda_{\theta}$ going towards negative values with increasing $p_{T}$ in both analysed colliding energies. And at $\sqrt{s} =$ 500 GeV we have also checked that the $\lambda_{\phi}$ parameter related to the azimuthal anisotropy is consistent with 0. Since the lower limit in the NRQCD calculations is unconstrained we do not draw conclusions from the comparison to data, and the new data points should help to constrain color-octet Long-Distance Matrix Elements for the model. The right panel of Fig.~\ref{fig:Jpsi500Polarization} shows $\lambda_{inv}$ parameters as a function of $p_{T}$ extracted in the HX and CS frames, shown as full and open symbols, respectively. $\lambda_{inv}$ has negative values and the results are in agreement between the frames.

\section{Summary}

In summary, STAR has measured the inclusive J/$\psi$ cross section and polarization in $p$+$p$ collisions at $\sqrt{s} =$ 200 and 500 GeV as a function of $p_{T}$. The measurements are compared to different model predictions of the J/$\psi$ production. The $p_{T}$ spectra are described well by the NRQCD calculations. The polarization parameter $\lambda_{\theta}$ at $\sqrt{s} =$ 200 GeV is consistent with the NLO$^{+}$ CSM. At both colliding energies $\lambda_{\theta}$ has a trend towards negative values with increasing $p_{T}$ in the HX frame. And at $\sqrt{s} =$ 500 GeV in the HX frame the $\lambda_{\phi}$ parameter is consistent with 0. Also, frame invariant parameters, $\lambda_{inv}$, agree in the HX and CS frames.
The first measurement of $\psi(2S) / J/\psi$ ratio in $p$+$p$ collisions at $\sqrt{s} =$ 500 GeV is  reported and compared with results from other experiments. No collision energy dependence is observed. A strong correlation is observed between relative J/$\psi$ yield and the event activity which is stronger than linear at higher multiplicities at $p_{T} >$ 4 GeV/$c$.

\vspace{-10pt}
\section*{Acknowledgements}

This publication was supported by the European social fund within the framework of realizing the project ,,Support of inter-sectoral mobility and quality enhancement of research teams at Czech Technical University in Prague'', CZ.1.07/2.3.00/30.0034. 

\section*{References}

\bibliographystyle{iopart-num} 
\bibliography{SQMbib}

\end{document}